\documentclass[10pt]{IEEEtran}
\usepackage{url}
\usepackage[utf8]{inputenc}
\usepackage[table,xcdraw]{xcolor}
\usepackage{amsmath}
\usepackage{amssymb}
\usepackage{xfrac}

\usepackage[singlespacing]{setspace} 
\usepackage[acronyms,nonumberlist,nopostdot,nomain,nogroupskip]{glossaries}
\usepackage{tablefootnote}
\usepackage{booktabs}
\usepackage{comment}

\usepackage{tabularx}
\usepackage{booktabs}

\usepackage{tikz}
\usepackage{pgfplots}
\pgfplotsset{compat=newest}
\pgfplotsset{plot coordinates/math parser=false}
\newlength\fheight
\newlength\fwidth
\usetikzlibrary{plotmarks,patterns,decorations.pathreplacing,backgrounds,calc,arrows,arrows.meta,spy,matrix}
\usepgfplotslibrary{patchplots,groupplots}
\usepackage{tikzscale}
\usepackage{hyperref}

\newif\ifexttikz
\exttikzfalse

\ifexttikz
	\usetikzlibrary{external}
\fi

\usepackage{multirow}
\usepackage[inline]{enumitem}
\usepackage{pgfplotstable} 
\usepackage{multirow}
\usetikzlibrary {patterns.meta}
\tikzset{
hatch size/.store in=\hatchsize,
hatch angle/.store in=\hatchangle,
hatch line width/.store in=\hatchlinewidth,
hatch size=5pt,
hatch angle=0pt,
hatch line width=.5pt,
}

\usepackage{subcaption}
\usepackage{caption}

\usepackage{mathtools}

\newacronym{ca}{CA}{Carrier Aggregation}
\newacronym{3gpp}{3GPP}{3rd Generation Partnership Project}
\newacronym{5g}{5G}{5th generation}
\newacronym{5gc}{5GC}{5G Core}
\newacronym{adc}{ADC}{Analog to Digital Converter}
\newacronym{afbw}{AFBW}{Average Fading Bandwidth}
\newacronym{aimd}{AIMD}{Additive Increase Multiplicative Decrease}
\newacronym{am}{AM}{Acknowledged Mode}
\newacronym{amc}{AMC}{Adaptive Modulation and Coding}
\newacronym{aoa}{AoA}{Angle of Arrival}
\newacronym{aod}{AoD}{Angle of Departure}
\newacronym{ap}{AP}{Access Point}
\newacronym{aqm}{AQM}{Active Queue Management}
\newacronym{awgn}{AGWN}{Additive White Gaussian Noise}
\newacronym{balia}{BALIA}{Balanced Link Adaptation}
\newacronym{bdp}{BDP}{Bandwidth-Delay Product}
\newacronym{ber}{BER}{Bit Error Rate}
\newacronym{bf}{BF}{Beamforming}
\newacronym{bwp}{BWP}{Bandwidth Part}
\newacronym{cad}{CAD}{Computer-Aided Design}
\newacronym{cc}{CC}{Congestion Control}
\newacronym{cdf}{CDF}{Cumulative Distribution Function}
\newacronym{cir}{CIR}{Channel Impulse Response}
\newacronym{cn}{CN}{Core Network}
\newacronym{cp}{CP}{Control Plane}
\newacronym{cqi}{CQI}{Channel Quality Information}
\newacronym{crs}{CRS}{Cell Reference Signal}
\newacronym{csirs}{CSI-RS}{Channel State Information - Reference Signal}
\newacronym{dc}{DC}{Dual Connectivity}
\newacronym{dce}{DCE}{Direct Code Execution}
\newacronym{dci}{DCI}{Downlink Control Information}
\newacronym{dl}{DL}{Downlink}
\newacronym{dmr}{DMR}{Deadline Miss Ratio}
\newacronym{dmrs}{DMRS}{DeModulation Reference Signal}
\newacronym{dray}{D-Ray}{Deterministic Ray}
\newacronym{e2e}{E2E}{End-to-End}
\newacronym{ecn}{ECN}{Explicit Congestion Notification}
\newacronym{edf}{EDF}{Earliest Deadline First}
\newacronym{enb}{eNB}{evolved Node Base}
\newacronym{endc}{EN-DC}{E-UTRAN-\gls{nr} \gls{dc}}
\newacronym{epc}{EPC}{Evolved Packet Core}
\newacronym{es}{ES}{Edge Server}
\newacronym{fdd}{FDD}{Frequency Division Duplexing}
\newacronym{fdma}{FDMA}{Frequency Division Multiple Access}
\newacronym{fray}{F-Ray}{Flashing Ray}
\newacronym{fs}{FS}{Fast Switching}
\newacronym{ftp}{FTP}{File Transfer Protocol}
\newacronym{gmm}{GMM}{Gaussian Mixture Model}
\newacronym{gnb}{gNB}{Next Generation Node Base}
\newacronym{harq}{HARQ}{Hybrid Automatic Repeat reQuest}
\newacronym{hetnet}{HetNet}{Heterogeneous Network}
\newacronym{hh}{HH}{Hard Handover}
\newacronym{hol}{HOL}{Head-of-Line}
\newacronym{hqf}{HQF}{Highest-quality-first}
\newacronym{ia}{IA}{Initial Access}
\newacronym{iab}{IAB}{Integrated Access and Backhaul}
\newacronym{imt}{IMT}{International Mobile Telecommunication}
\newacronym{inf}{InF}{Indoor Factory}
\newacronym{inr}{INR}{Interference to Noise Ratio}
\newacronym{iot}{IoT}{Internet of Things}
\newacronym{ked}{KED}{Knife-Edge Diffraction}
\newacronym{kpi}{KPI}{Key Performance Indicator}
\newacronym{lcf}{LCF}{Level Crossing Frequency}
\newacronym{lcr}{LCR}{Level Crossing Rate}
\newacronym{los}{LoS}{Line-of-Sight}
\newacronym{lte}{LTE}{Long Term Evolution}
\newacronym{ltemtp}{LTE-M}{LTE-MTC [Machine Type Communication]}
\newacronym{m2m}{M2M}{Machine to Machine}
\newacronym{mac}{MAC}{Medium Access Control}
\newacronym{mc}{MC}{Multi-Connectivity}
\newacronym{mcs}{MCS}{Modulation and Coding Scheme}
\newacronym{mec}{MEC}{Mobile Edge Cloud}
\newacronym{mi}{MI}{Mutual Information}
\newacronym{mib}{MIB}{Master Information Block}
\newacronym{mimo}{MIMO}{Multiple Input Multiple Output}
\newacronym{m-mimo}{m-MIMO}{massive-MIMO}
\newacronym{mlr}{MLR}{Maximum-local-rate}
\newacronym{mmwave}{mmWave}{millimeter wave}
\newacronym{moi}{MoI}{Method of Images}
\newacronym{mpc}{MPC}{Multi Path Component}
\newacronym{mptcp}{MPTCP}{Multipath TCP}
\newacronym{mr}{MR}{Maximum Rate}
\newacronym{mrdc}{MR-DC}{Multi \gls{rat} \gls{dc}}
\newacronym{mss}{MSS}{Maximum Segment Size}
\newacronym{mtd}{MTD}{Machine-Type Device}
\newacronym{mtu}{MTU}{Maximum Transmission Unit}
\newacronym{nfv}{NFV}{Network Function Virtualization}
\newacronym{nist}{NIST}{National Institute of Standards and Technology}
\newacronym{nlos}{NLoS}{Non-Line-of-Sight}
\newacronym{nr}{NR}{New Radio}
\newacronym{nrmse}{NRMSE}{Normalized Root Mean Square Error}
\newacronym{nsa}{NSA}{Non Stand Alone}
\newacronym{o2i}{O2I}{Outdoor-to-Indoor}
\newacronym{ofdm}{OFDM}{Orthogonal Frequency Division Multiplexing}
\newacronym{pa}{PA}{Position-aware}
\newacronym{prr}{PRR}{Packet Reception Ratio}
\newacronym{pbch}{PBCH}{Physical Broadcast Channel}
\newacronym{pdcch}{PDCCH}{Physical Downlonk Control Channel}
\newacronym{pdcp}{PDCP}{Packet Data Convergence Protocol}
\newacronym{pdsch}{PDSCH}{Physical Downlink Shared Channel}
\newacronym{pdu}{PDU}{Packet Data Unit}
\newacronym{per}{PER}{Packet Error Rate}
\newacronym{pf}{PF}{Proportional Fair}
\newacronym{pgw}{PGW}{Packet Gateway}
\newacronym{phy}{PHY}{Physical}
\newacronym{pl}{PL}{Path Loss}
\newacronym{ppp}{PPP}{Poisson Point Process}
\newacronym{prb}{PRB}{Physical Resource Block}
\newacronym{pss}{PSS}{Primary Synchronization Signal}
\newacronym{pucch}{PUCCH}{Physical Uplink Control Channel}
\newacronym{pusch}{PUSCH}{Physical Uplink Shared Channel}
\newacronym{qam}{QAM}{Quadrature Amplitude Modulation}
\newacronym{qd}{QD}{Quasi Deterministic}
\newacronym{rach}{RACH}{Random Access Channel}
\newacronym{ran}{RAN}{Radio Access Network}
\newacronym[firstplural=Radio Access Technologies (RATs)]{rat}{RAT}{Radio Access Technology}
\newacronym{red}{RED}{Random Early Detection}
\newacronym{rf}{RF}{Radio Frequency}
\newacronym{rlc}{RLC}{Radio Link Control}
\newacronym{rlf}{RLF}{Radio Link Failure}
\newacronym{rr}{RR}{Round Robin}
\newacronym{rray}{R-Ray}{Random Ray}
\newacronym{rrc}{RRC}{Radio Resource Control}
\newacronym{rrm}{RRM}{Radio Resource Management}
\newacronym{rs}{RS}{Remote Server}
\newacronym{rsrp}{RSRP}{Reference Signal Received Power}
\newacronym{rsrq}{RSRQ}{Reference Signal Received Quality}
\newacronym{rss}{RSS}{Received Signal Strength}
\newacronym{rssi}{RSSI}{Received Signal Strength Indicator}
\newacronym{rt}{RT}{Ray Tracer}
\newacronym{rtt}{RTT}{Round Trip Time}
\newacronym{rw}{RW}{Receive Window}
\newacronym{rx}{RX}{Receiver}
\newacronym{sa}{SA}{standalone}
\newacronym{sack}{SACK}{Selective Acknowledgment}
\newacronym{sap}{SAP}{Service Access Point}
\newacronym{sch}{SCH}{Secondary Cell Handover}
\newacronym{scm}{SCM}{Spatial Channel Model}
\newacronym{scoot}{SCOOT}{Split Cycle Offset Optimization Technique}
\newacronym{sdma}{SDMA}{Spatial Division Multiple Access}
\newacronym{sf}{SF}{Shadow Fading}
\newacronym{si}{SI}{Study Item}
\newacronym{sib}{SIB}{Secondary Information Block}
\newacronym{sinr}{SINR}{Signal-to-Interference-plus-Noise Ratio}
\newacronym{sir}{SIR}{Signal-to-Interference Ratio}
\newacronym{sm}{SM}{Saturation Mode}
\newacronym{snr}{SNR}{Signal-to-Noise Ratio}
\newacronym{son}{SON}{Self-Organizing Network}
\newacronym{srs}{SRS}{Sounding Reference Signal}
\newacronym{ss}{SS}{Synchronization Signal}
\newacronym{sss}{SSS}{Secondary Synchronization Signal}
\newacronym{sta}{STA}{Station}
\newacronym{tb}{TB}{Transport Block}
\newacronym{tcp}{TCP}{Transmission Control Protocol}
\newacronym{udp}{UDP}{User Datagram Protocol}
\newacronym{tdd}{TDD}{Time Division Duplexing}
\newacronym{tdma}{TDMA}{Time Division Multiple Access}
\newacronym{tfl}{TfL}{Transport for London}
\newacronym{tgad}{TGad}{Task Group ad}
\newacronym{tgay}{TGay}{Task Group ay}
\newacronym{tm}{TM}{Transparent Mode}
\newacronym{trp}{TRP}{Transmitter Receiver Pair}
\newacronym{tti}{TTI}{Transmission Time Interval}
\newacronym{ttt}{TTT}{Time-to-Trigger}
\newacronym{tx}{TX}{Transmitter}
\newacronym{ue}{UE}{User Equipment}
\newacronym{ul}{UL}{Uplink}
\newacronym{um}{UM}{Unacknowledged Mode}
\newacronym{uma}{UMa}{Urban Macro}
\newacronym{uml}{UML}{Unified Modeling Language}
\newacronym{utc}{UTC}{Urban Traffic Control}
\newacronym{vm}{VM}{Virtual Machine}
\newacronym{wbf}{WBF}{Wired Bias Function}
\newacronym{wf}{WF}{Wired-first}
\newacronym{wifi}{Wi-Fi}{Wireless Fidelity}
\newacronym{wigig}{WiGig}{Wireless Gigabit}
\newacronym{wlan}{WLAN}{Wireless Local Area Network}
\newacronym{xpr}{XPR}{Cross Polarization Ratio}
\newacronym{fr2}{FR2}{Frequency Range 2}
\newacronym{nbiot}{NB-IoT}{Narrowband-IoT}
\newacronym{cps}{CPS}{Cyber-Physical production System}
\newacronym{iiot}{IIoT}{Industrial Internet of Things}
\newacronym{agv}{AGV}{Autonomous Ground Vehicle}
\newacronym{uav}{UAV}{Unmanned Autonomous Vehicle}
\newacronym{amr}{AMR}{Autonomous Mobile Robots}
\newacronym{wsn}{WSN}{Wireless Sensor Network}
\newacronym{embb}{eMBB}{enhanced Mobile Broadband}
\newacronym{urllc}{URLLC}{Ultra-Reliable Low-Latency Communications}

\tikzstyle{startstop} = [rectangle, rounded corners, minimum width=2cm, minimum height=0.5cm,text centered, draw=black]
\tikzstyle{io} = [trapezium, trapezium left angle=70, trapezium right angle=110, minimum width=3cm, minimum height=1cm, text centered, draw=black]
\tikzstyle{process} = [rectangle, minimum width=2cm, minimum height=0.5cm, text centered, draw=black, alignb=center]
\tikzstyle{decision} = [ellipse, minimum width=2cm, minimum height=1cm, text centered, draw=black]
\tikzstyle{arrow} = [thick,<->,>=stealth]
\tikzstyle{line} = [thick,>=stealth]
\tikzstyle{darrow} = [thick,<->,>=stealth,dashed]
\tikzstyle{sarrow} = [thick,->,>=stealth]
\tikzstyle{larrow} = [line width=0.1mm,dashdotted,->,>=stealth]

\makeatletter
\def\grd@save@target#1{%
  \def\grd@target{#1}}
\def\grd@save@start#1{%
  \def\grd@start{#1}}
\tikzset{
  grid with coordinates/.style={
    to path={%
      \pgfextra{%
        \edef\grd@@target{(\tikztotarget)}%
        \tikz@scan@one@point\grd@save@target\grd@@target\relax
        \edef\grd@@start{(\tikztostart)}%
        \tikz@scan@one@point\grd@save@start\grd@@start\relax
        \draw[minor help lines] (\tikztostart) grid (\tikztotarget);
        \draw[major help lines] (\tikztostart) grid (\tikztotarget);
        \grd@start
        \pgfmathsetmacro{\grd@xa}{\the\pgf@x/1cm}
        \pgfmathsetmacro{\grd@ya}{\the\pgf@y/1cm}
        \grd@target
        \pgfmathsetmacro{\grd@xb}{\the\pgf@x/1cm}
        \pgfmathsetmacro{\grd@yb}{\the\pgf@y/1cm}
        \pgfmathsetmacro{\grd@xc}{\grd@xa + \pgfkeysvalueof{/tikz/grid with coordinates/major step x}}
        \pgfmathsetmacro{\grd@yc}{\grd@ya + \pgfkeysvalueof{/tikz/grid with coordinates/major step y}}
        \foreach \x in {\grd@xa,\grd@xc,...,\grd@xb}
        \node[anchor=north] at (\x,\grd@ya) {\pgfmathprintnumber{\x}};
        \foreach \y in {\grd@ya,\grd@yc,...,\grd@yb}
        \node[anchor=east] at (\grd@xa,\y) {\pgfmathprintnumber{\y}};
      }
    }
  },
  minor help lines/.style={
    help lines,
    gray,
    line cap =round,
    xstep=\pgfkeysvalueof{/tikz/grid with coordinates/minor step x},
    ystep=\pgfkeysvalueof{/tikz/grid with coordinates/minor step y}
  },
  major help lines/.style={
    help lines,
    line cap =round,
    line width=\pgfkeysvalueof{/tikz/grid with coordinates/major line width},
    xstep=\pgfkeysvalueof{/tikz/grid with coordinates/major step x},
    ystep=\pgfkeysvalueof{/tikz/grid with coordinates/major step y}
  },
  grid with coordinates/.cd,
  minor step x/.initial=.5,
  minor step y/.initial=.2,
  major step x/.initial=1,
  major step y/.initial=1,
  major line width/.initial=1pt,
}
\makeatother

\definecolor{desireRed}{RGB}{230,57,60}%
\definecolor{darkPurple}{RGB}{59,31,43}%
\definecolor{springGreen}{RGB}{37,223,145}%
\definecolor{queenBlue}{RGB}{69,123,157}%
\definecolor{spaceCadet}{RGB}{29,53,87}%

\definecolor{primaryColor}{HTML}{F06449}
\definecolor{secondaryColor}{HTML}{5BC3EB}
\definecolor{tertiaryColor}{HTML}{36382E}

\usepackage[font=footnotesize]{caption}
\usepackage[font=footnotesize]{subcaption}

\makeglossaries

\begin{document}
\bstctlcite{IEEEexample:BSTcontrol}

\title{5G NR-Light at Millimeter Waves: \\ Design Guidelines for Mid-Market IoT Use Cases}

\author{\IEEEauthorblockN{ Matteo Pagin$^*$, Tommaso Zugno$^{*\circ}$, Marco Giordani$^*$, Louis-Adrien Dufrene$^\dagger$, Quentin Lampin$^\dagger$, Michele Zorzi$^*$}
\IEEEauthorblockA{\\
$^*$ University of Padova, Italy (email: \texttt{\{name.surname\}@dei.unipd.it})\\
$^\dagger$ Orange Labs (email: \texttt{\{name.surname\}@orange.com})\\
\thanks{$^{\circ }$ Tommaso Zugno is now with Huawei Technologies Duesseldorf GmbH.}
}}



\maketitle

\begin{abstract}
  5th generation (5G) systems have been designed with three main objectives in mind: increasing throughput, reducing latency, and enabling reliable communications.
  To meet these (often conflicting) constraints, the 3GPP released a set of specifications for 5G NR, one of the main innovations being the support for communications in the millimeter wave (mmWave) bands.
  However, how to implement lower complexity, energy efficient, mid-market Internet of Things (IoT) applications is still an on-going investigation, currently led by the 3GPP which is extending the NR standard with NR-Light specifications to support devices with reduced capabilities (REDCAP).
  While REDCAP devices may also operate at mmWaves to improve the network performance, hardware/software simplifications are needed to support balanced and mixed requirements compared to 5G NR systems.
In this context, the contributions of this paper are threefold. 
  First, we present some NR-Light use cases for which the support of the mmWave bands is desirable.
  Second, we describe how 5G NR can be simplified to achieve~\mbox{NR-Light} requirements and expectations. 
  Finally, we evaluate via simulation the performance of NR-Light devices operating at mmWaves in an industrial IoT setup,
in terms of cost and complexity, throughput, and latency.  
\end{abstract}


\begin{picture}(0,0)(-95,-430)
\put(0,-10){\footnotesize This paper has been submitted to IEEE for publication. Copyright may change without notice.}
\end{picture}

\glsresetall

\begin{figure*}[t!]
  \centering
  \includegraphics[width=0.9\textwidth]{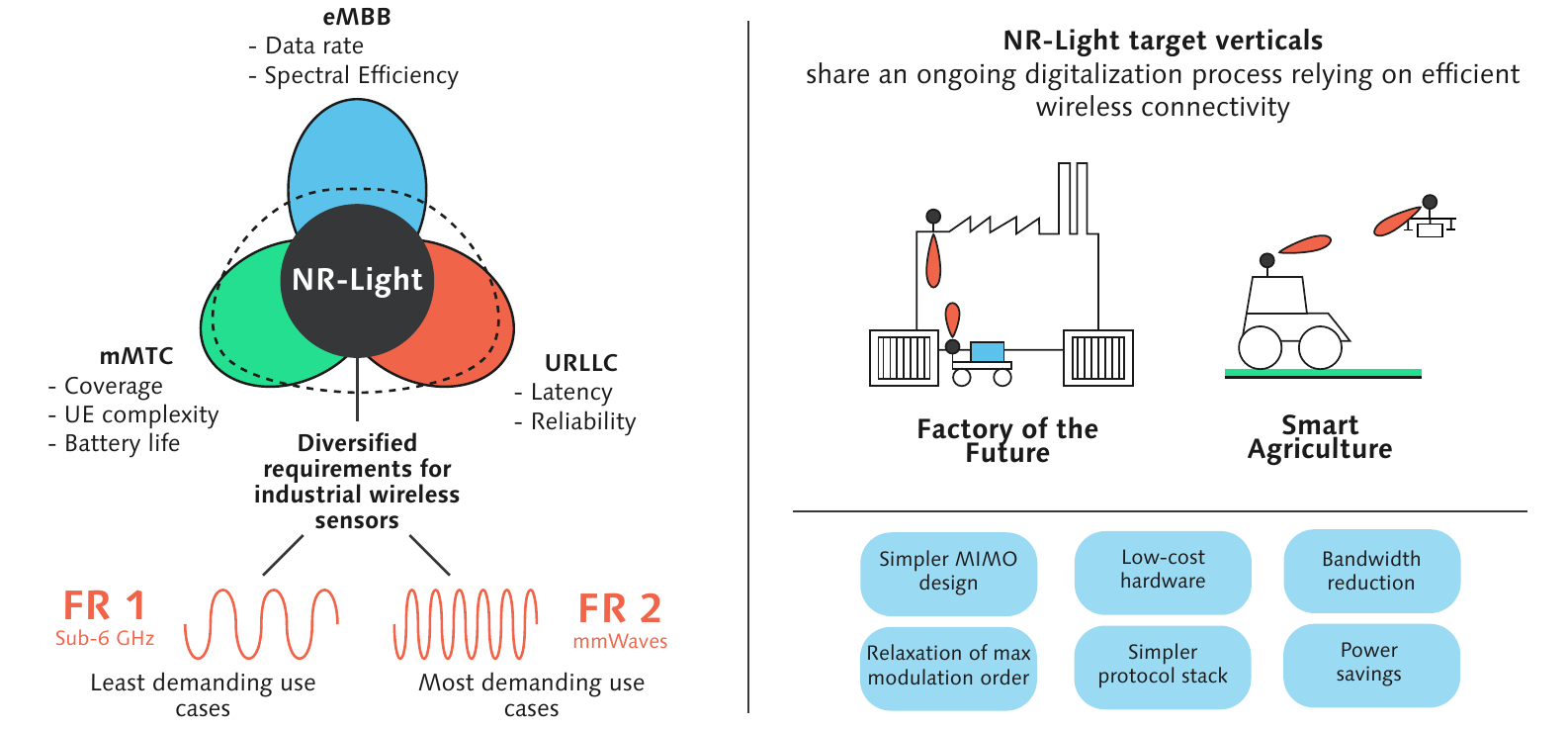}
  \caption{Overview of NR-Light use cases and target Key Performance Indicator (KPI) requirements.}
  \label{Fig:overview}
\end{figure*}

\section{Introduction}
\label{sec:intro}
The grand objective of \gls{5g} systems is to support three generic services with vastly heterogeneous requirements: enhanced mobile broadband (eMBB), massive machine-type communication (mMTC), and ultra-reliable low-latency communication (URLLC).
5G guarantees very high data rates (up to 20 Gbps in ideal conditions), ultra-low latency (around 1 ms), and a 10$\times$ increase in energy efficiency with respect to previous wireless generations.
To meet those requirements, the \gls{3gpp} has released a set of specifications for NR~\cite{38300}
which include, besides an updated/flexible radio access and core network design, the support for communications in the  \gls{mmwave} range up to 52.6 GHz for Release 15. 
On one hand, the vast amount of available spectrum at mmWaves makes it possible to achieve multi-Gbps transmission speeds, while also improving security and privacy thanks to directional transmissions~\cite{rappaport2013millimeter}.
On the other hand, exploiting \glspl{mmwave} is challenging for mid-market \gls{iot} devices, such as wearables or industrial~sensors, which are subject to cost, complexity, and battery lifetime constraints.

To address the above market use cases, the \gls{3gpp} is extending NR specifications to support a simpler and lighter version of NR, referred to as NR-Light (or ``reduced-capability (REDCAP) NR devices'' in 3GPP parlance)~\cite{varsier20215g}.
NR-Light needs to satisfy higher data rate, improved reliability, and lower latency than current LTE-based LTE-MTC and \gls{nbiot} technologies for IoT services, while guaranteeing lower cost/complexity, longer battery life, and wider coverage than 5G NR solutions.
Among other features, NR-Light is expected to support network operation in both FR1 (between 410 MHz and 7125 MHz) and FR2 (between 24.25 and 52.6 GHz), i.e., in the lower part of the \gls{mmwave} spectrum, to improve the network performance.
While being attractive for private networks, e.g., to support high-end applications in the industrial domain, 
\glspl{mmwave} raise many challenges for \gls{iot} use cases, including how to meet the low-cost/low-power requirements of NR-Light while providing sufficient performance levels. 
Recently, the 3GPP has proposed a set of simplification of the 5G NR standard to support lower-complexity energy-efficient NR-Light devices at mmWaves~\cite{38875}, which are often cumbersome to retrieve and read, thus complicating research in this field. 
Also, 3GPP documents often miss adequate numerical validation of the performance of those simplifications. 

To fill this gap, in this paper we shed light on some IoT use cases that will benefit the most from NR-Light at \glspl{mmwave}, and characterize their \glspl{kpi}.
Then, we review possible hardware/software simplifications of the 5G NR standard to fulfill NR-Light requirements in mmWave bands while reducing energy consumption and costs. These include (i) narrower bandwidth for devices, (ii) simplified air interface procedures, protocol stack, and antenna configuration, and (iii) enhanced device power saving features, as summarized in Fig.~\ref{Fig:overview}.
Finally, we provide guidelines, based on numerical simulations, towards an efficient set of simplifications for NR-Light devices, referred to as ``NR-L-Mid,'' in an indoor factory scenario (although we do not preclude other simplifications from being adopted too). 
Our simulations reveal that demanding Industrial IoT applications involving video streaming and periodic reporting of sensor readings can be advantageously supported by NR-L-Mid.

\section{IoT Use Cases for NR-Light at mmWaves}
\label{use_cases}
Originally, IoT use cases are characterized by three fundamental prerequisites, namely the support for (i) long-range links (in the orders of kilometers), (ii) bursty traffic with loose performance requirements, and (iii) low-energy-consuming low-complexity architectures and network functionalities~\cite{zanella2014internet}.
On the other hand, future IoT use cases for NR-Light expose bolder requirements in terms of data rate, reliability, and latency, for which the large bandwidth available at \gls{mmwave} is an attractive option.
Notably, \gls{mmwave} bands are cheaper with respect to more common sub-6~GHz bands (FR1), allowing operators to buy large chunks of continuous spectrum with limited investments.
Also, the foreseen directionality of \gls{mmwave} communications can provide limited interference and enable higher spectral efficiency through spatial diversity.
Additionally, the limited propagation range and the inability to penetrate through walls make mmWaves suitable for the realization of private networks, such as in factories, where private cells can be placed in different areas without interfering with other stations and without propagating outside buildings.

\begin{figure*}[t!]
  \centering
  \captionof{table}{Communication requirements for NR-Light use cases (``factory of the future'' and ``smart agriculture'' in light and dark grey, respectively)~\cite{22804}.}
  \label{tab:req}
  \includegraphics[width=0.99\textwidth]{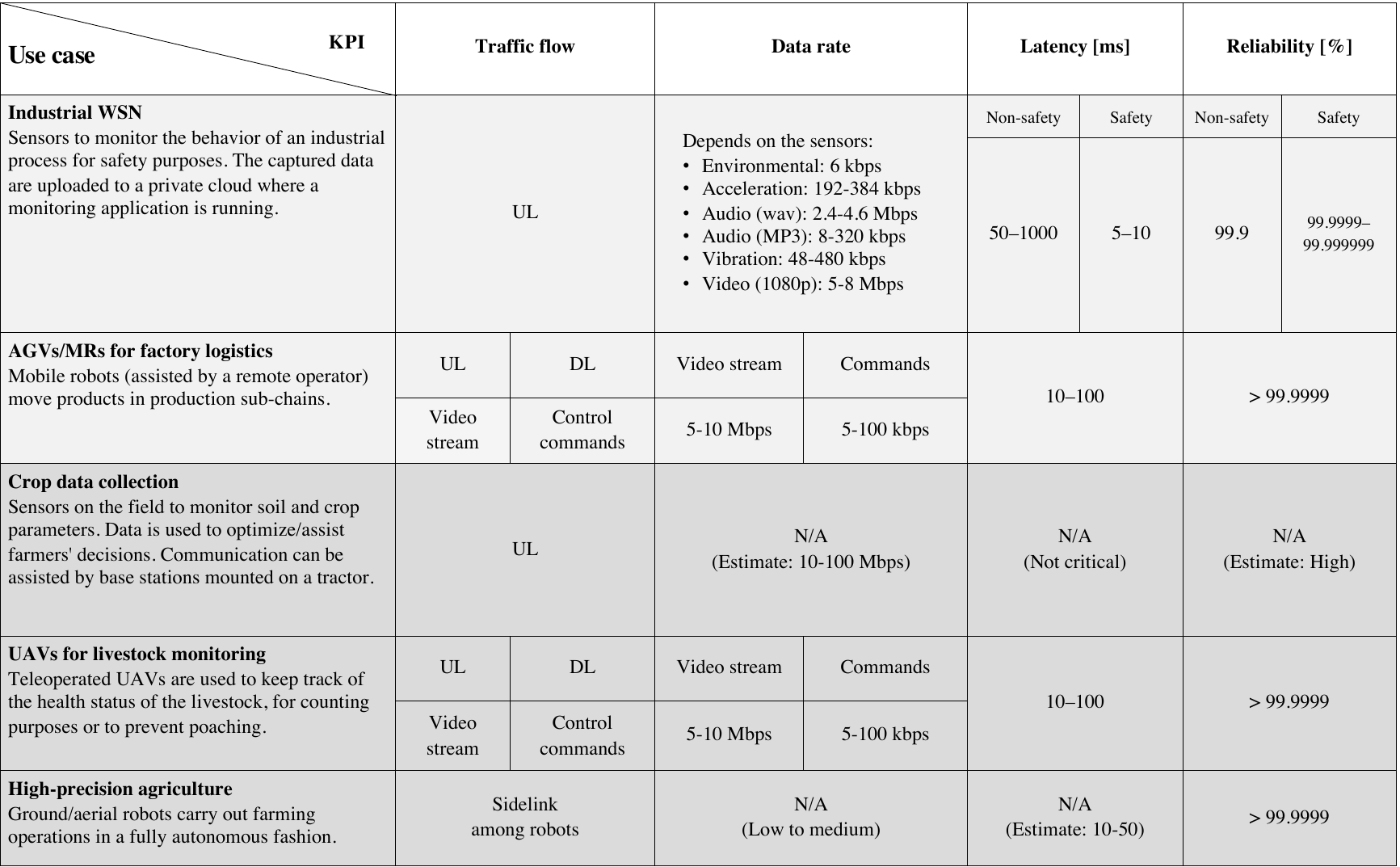}
\end{figure*}

In this section we present two representative use cases for NR-Light, namely factory of the future (Sec.~\ref{sub:foth}) and smart agriculture (Sec.~\ref{sub:sa}), for which \glspl{mmwave} are desirable, while the relative \glspl{kpi} (based on 3GPP specifications in~\cite{22804}) are summarized in Table~\ref{tab:req}.

\subsection{Factory of the Future}
\label{sub:foth}
Increasingly volatile/globalized markets are opening the way towards \gls{iiot} to support ubiquitous connectivity and powerful computing infrastructures in the industrial sector.
Among the many use cases in this ecosystem
we can identify some common characteristics~\cite{8502649}.
For example, signal propagation in factories may be very different from a typical cellular deployment, because these scenarios feature a much richer scattering environment and the presence of many blockers. 
Also, networks may be private and built to cover a limited area.
Some critical use cases may have stringent requirements in terms of reliability and privacy.
Finally, a longer lifetime of the communication infrastructure may be required compared to other wireless systems.
\medskip

\emph{The role of NR-Light.} NR-Light at \glspl{mmwave} can play a key role to support the digitalization process of future factories. 

For example, non-time-critical \glspl{agv} and Mobile Robots (MRs) are expected to improve the efficiency of logistics and warehousing, and to automate industrial operations~\cite{boban2021predictive}. 
The traffic flow consists of video feeds from various sensors, either periodic (e.g., in case of teleoperated remote control) or aperiodic (e.g., in case of fully-automated robots).
In both cases, the latency requirements go from 10 to 100~ms, and the packet error rate should be below $10^{-5}$~\cite{cuozzo2022enabling}. 
Therefore, the vast bandwidth available at \glspl{mmwave} can support the exchange of sensors' data at high speed.
Moreover, \glspl{agv}/MRs are beneficial only if affordable: in~\cite{9247159}, the authors estimate that the cost of mobile robots should be lower than $1000$ USD/year to make this solution economically viable. In this context, NR-Light modules will be cheaper compared to 5G NR modems, and will then be preferable.


Another promising paradigm for the factories of the future is the adoption of \glspl{wsn}, i.e., the deployment of a large number of sensors to monitor the state of an industrial process. As such, the 
directionality of \gls{mmwave} links could reduce interference among different sensors, and maximize the spatial reuse of radio resources.
Furthermore, the form factor of NR-Light modules shall be smaller with respect to 5G NR hardware, therefore limiting the size/cost of the sensors.
The data generated by the sensors are then collected and exchanged at a centralized controller which can run specialized algorithms to detect anomalies and/or optimize the inputs.
In this case, the traffic pattern depends on the types of measurements: low-bandwidth (e.g., a few kbps for environmental signals) or high-bandwidth (e.g., in the order of Mbps for video/images) streams might be transmitted. In particular, each sensor may generate a small amount of (bursty) traffic, but the aggregate data rates might be high.
In this perspective, the NR-Light architecture guarantees flexibility, and accommodates the different traffic patterns involved.


\subsection{Smart Agriculture}
\label{sub:sa}
The growth of human population, the consequent rising demand for agricultural products, and the need to protect the environment, triggered new innovations in the agricultural sector. Specifically, digital technologies will play a crucial role in maximizing the production chain output, optimizing the use of the available land resources, minimizing both inputs (water, fertilizers, pesticides) and waste, and reducing CO$_2$ emissions~\cite{8066090}.
For example, the pervasive use of \glspl{uav} may facilitate harvesting and pest monitoring of the livestock, while robots using vision positioning systems may identify and locate the fruit to harvest~\cite{giordani2021non}.
This paradigm, typically referred to as ``smart agriculture,'' is being promoted by various international organizations, such as the agricultural European Innovation Partnership (EIP-AGRI).

In general, the agriculture environment exhibits favorable channel conditions (as the signal typically propagates in remote areas without buildings/obstacles), but relatively vast coverage areas are needed.
Moreover, network operations rely on limited or even absent communication infrastructures and/or power grid, and may require renewable energy supplies, thus further motivating the use of low-power devices.
\medskip

\emph{The role of NR-Light.}
NR-Light provides significant power enhancements with respect to the NR standard, and thus is well suited for agriculture scenarios in which devices are battery powered and the energy used by the radio module should be minimized.
At the same time, NR-Light at \glspl{mmwave}  promotes better coordination among swarms of robots, and supports offloading of data with low latency.
Finally,  beamforming, as typically established at \glspl{mmwave},  can provide a valid alternative to estimate the relative position of agriculture robots (e.g., in the crop field) when GPS sensors are not available, for example to reduce both costs and power consumption.

\section{Key Technology Enablers and Simplifications \\ for NR-Light at MmWaves}
\label{sec:tech}
NR-Light targets vertical domains which, in most cases, involve more powerful devices than in traditional \gls{iot} but, at the same time, cheaper and less power hungry than high-end 5G NR terminals. 
Furthermore, 5G NR operations in the \gls{mmwave} bands look incompatible with these prerequisites.
Along these lines, in this section we present possible simplifications of the 5G NR standard to support NR-Light at mmWaves, based on 3GPP specifications and current trends~\cite{38875}.

\subsection{Simpler \gls{mimo} design}
\label{sec:tech_mimo}

The main feature of \gls{mmwave} systems is the realization of \gls{m-mimo} to overcome the severe propagation loss experienced at high frequencies.
However, a typical m-MIMO architecture requires several hardware components, thus consuming substantial energy.
In turn, NR-Light use cases, like those in Sec.~\ref{use_cases}, require low power consumption and reduced complexity, therefore a simplification of the radio front end might be desirable.
For instance, unlike legacy mmWave terminals that may incorporate even thousands of antennas, NR-Light devices may satisfy looser performance requirements at much smaller form factors, potentially including a relaxation of the number of MIMO layers.

At the same time, the research community should investigate which beamforming architecture 
would better support NR-Light use cases.
Digital and hybrid architectures provide the best communication performance and flexibility, enabling the transceiver to direct beams in several directions simultaneously, despite involving more power-hungry blocks.
While novel proposals suggest the usage of low-resolution \glspl{adc} in the RF chain~\cite{zhang2018low}, an analog architecture seems the most suitable choice for NR-Light devices to minimize power consumption.



\subsection{Low-Cost Hardware Components}
A critical property of \gls{mmwave} systems is phase noise, which results in random oscillations in the phase of the signal, thus reducing the achievable spectral efficiency.
This problem can be mitigated by using high-precision oscillators, which are expensive but able to produce a low-noise carrier signal.
However, it may not be possible to apply this solution to NR-Light devices, since the cost may be too high.
A more cost-effective solution is the configuration of 5G-NR-specific Phase Tracking Reference Signals (PTRSs)~\cite{dahlman20205g} to track and compensate phase noise variations within the slot, and the adoption of a higher spacing between the subcarriers to avoid inter-carrier interference.

Another possible approach to decrease the cost of NR-Light devices consists of using lens-based antenna arrays.
In analog architectures, one third of the total cost of the antenna array is due to phase shifters~\cite{8613274}, and this term is proportional to the number of antenna elements in the array.
Lens-based antennas represent a cheaper alternative to regular arrays, where beamforming is achieved through an electromagnetic lens, without the need for phase shifters.

Another promising research aspect to explore is how to reduce the cost of physical NR-Light hardware by cloud-based implementation of the \gls{ran}, thus realizing NR-Light functionalities via software.

\subsection{Bandwidth Reduction}
The high power consumption of the mmWave radio is due to its \gls{rf} components, whose power consumption grows linearly with the system bandwidth~\cite{abbas2017millimeter}. NR Release 15 devices support a bandwidth of up to 400 MHz per carrier in FR2, which can be further increased by means of \gls{ca}. In turn, the loose network requirements of NR-Light use cases could make it theoretically possible to reduce the supported bandwidth, thus minimizing the power consumption. Moreover, bandwidth reduction enables the employment of cheaper \glspl{adc}, since the required sampling rate is reduced. For example, reducing the bandwidth from 200 to 50~MHz saves up to 23.5\% of the cost of the radio module~\cite{38875}.
For these reasons, the 3GPP envisions a bandwidth ranging from 50 to 100 MHz for NR-Light devices operating in the \gls{mmwave} bands.

\subsection{Relaxation of the Maximum Modulation Order}
\label{ssub:msc}
The NR standard supports an efficient \gls{amc} mechanism, which permits to adjust the modulation order and the coding rate used by the transmitter depending on the channel quality. The maximum modulation order supported by NR mmWave devices is 8~\cite[Sec. 5.1.3]{38214}.
For NR-Light, the 3GPP suggests to limit this value to 4, hence decreasing the cost of hardware components (e.g., relaxing the maximum modulation order from 6 to 4 saves up to 5.6\% of the total cost) and reducing the processing time required to modulate/demodulate the~signals~\cite{38875}.

\subsection{Simplification of the Protocol Stack} 
5G NR's main novelty is flexibility: it supports multiple numerologies for the \gls{ofdm} waveform to accommodate diverse service requirements~\cite{polese20183gpp}. For NR-Light, flexibility can be extended to the whole protocol stack by configuring a scalable structure compared to the mandatory features of Release 15 NR, e.g., a reduced maximum transport block size, a relaxed physical data channel, a reduced and simplified measurement and reporting mechanism. This improved flexibility is theoretically forward compatible with 5G's slicing paradigm for adapting the protocol stack to the device requirements and transmission~profiles.

NR-Light can further promote energy efficiency by rethinking legacy NR protocol implementations.
Beam management mechanisms, for example, are mandatory in NR to ensure that the end devices are properly aligned while communicating~\cite{giordani2018tutorial}. Existing beam searching techniques, however, require continuous exchange of control signals, thereby increasing the power consumption and computation requirements of the device. On the other hand, in many NR-Light use cases, nodes do not move, thus making it possible to decrease the periodicity at which beam management signals are broadcast (or, equivalently, reduce the beam sweep space).

\begin{figure*}[t!]
  \centering
  \setlength\fwidth{0.7\textwidth}
  \setlength\fheight{0.25\textwidth}
  \begin{tikzpicture}

    \pgfplotsset{every tick label/.append style={font=\footnotesize}}

    \pgfplotsset{
        tick label style={font=\footnotesize},
        label style={font=\footnotesize},
    }

    \pgfplotsset{
        legend image code/.code={
            \draw [#1] (0cm, -0.1cm) rectangle (0.4cm,0.4cm);
        },
    }


     \begin{axis}[
        ybar,
        bar shift=0pt,
        bar width=1.1,
        width=\fwidth,
        height=\fheight,
        at={(0\fwidth,0\fheight)},
        ylabel={Latency [ms]},
        axis y line*=right,
        scale only axis,
        ymajorticks,
        xmajorgrids,
        ymajorgrids,
        xmin=-3, xmax=15,
        ymin=0, ymax=28,
        ytick style={color=white!15!black},
        xtick={-2, 0.8, 2.2, 4.8, 6.2, 8.8, 10.2, 12.8, 14.2},
        xticklabels={NR-L-Low, 100, 200, 16, 28, 4, 16, 18, 23},
        xtick pos=left
    ]
    \addplot [preaction={fill, white}, pattern={north east lines}, very thin] coordinates {
        (-2, 24.926070769672)
        };
    \addplot [preaction={fill, white},  pattern={north east lines}, very thin] coordinates {
        (0.8, 24.5421820832409)
        (2.2, 22.6181270233416)
         };

    \addplot [preaction={fill, white}, pattern={north east lines}, very thin] coordinates {
        (4.8, 25.2128156992463)
        (6.2, 25.4587916938308)
        };

    \addplot [preaction={fill, white}, very thin, pattern={north east lines},,
    pattern color=black] coordinates {
        (8.8, 15.5960420311129)
        (10.2, 10.445907741513)
        };

    \addplot [very thin, preaction={fill, white}, pattern={north east lines},] coordinates {
        (12.8, 14.5937116577583)
        (14.2, 10.6157863955618)
        };
    \end{axis}

    
    \begin{axis}[
        ybar,
        bar shift=0pt,
        bar width=0.5,
        hide x axis,
        axis y line*=left,
        width=\fwidth,
        height=\fheight,
        at={(0\fwidth,0\fheight)},
        ylabel={Trhoughout [Mbps]},
        scale only axis,
        xmajorticks=false,
        ymajorticks,
        ymajorgrids,
        xmin=-3, xmax=15,
        ymin=0, ymax=12,
        ytick style={color=white!15!black},
    ]
    \addplot [preaction={fill, primaryColor}, very thin] coordinates {
        (-2, 8.24581770076594)
        };

    \addplot [preaction={fill, primaryColor}, very thin] coordinates {
        (0.8, 8.25002439300989)
        (2.2, 7.86459514715495)
    };

    \addplot [preaction={fill, primaryColor}, very thin] coordinates {
        (4.8, 8.16142598635792)
        (6.2, 8.43446579178086)
        };

    \addplot [preaction={fill, primaryColor}, very thin] coordinates {
        (8.8, 9.30121433057365)
        (10.2, 9.86204979818936)
        };

    \addplot [preaction={fill, primaryColor}, very thin] coordinates {
        (12.8, 9.37257317512032)
        (14.2, 9.98595290026055)
        };

    \end{axis}

    \begin{axis}[%
        ybar,
        hide y axis,
        area legend,
        width=\fwidth,
        height=\fheight,
        at={(0\fwidth,0\fheight)},
        scale only axis,
        xmin=-3, xmax=15,
        ymin=0, ymax=28,
        xtick={-2,1.5, 5.5, 9.5, 13.5},
        xtick style={draw=none},
        xticklabels={(Baseline),BW [MHz], Max MCS index, Number antennas, Max TX power [dBm]},
        xticklabel style={yshift=-0.3cm},
        legend style={at={(0.5, 0.95)}, font=\footnotesize, anchor=south, draw=white!15!black, /tikz/every even column/.append style={column sep=0.3cm}},
        axis line style={-},
        legend columns=2
    ]

    \addplot [preaction={fill, primaryColor}, very thin]
    coordinates {
            (0,0)
        };
    \addlegendentry{Throughput};

    \addplot [preaction={fill, white}, pattern={north east lines}, very thin]
    coordinates {
            (0,0)
        };
    \addlegendentry{Latency};

    \end{axis}

\end{tikzpicture}
  \caption{Average throughput (for video-stream) and latency (for data-sensor) of NR-Light devices in an InF scenario, as a function of the bandwidth ($\{100, 200 \}$ MHz), the maximum MCS index ($\{16, 28 \}$), the number of antenna elements ($\{4, 16 \}$), and the maximum transmission power ($\{18, 23 \}$ dBm).
    We consider a baseline configuration called ``NR-L-Low'' with a bandwidth of 50 MHz, a maximum modulation order of 2, one antenna element, and a maximum transmission power of 13 dBm.}
  \label{fig:perf-barplot}
\end{figure*}
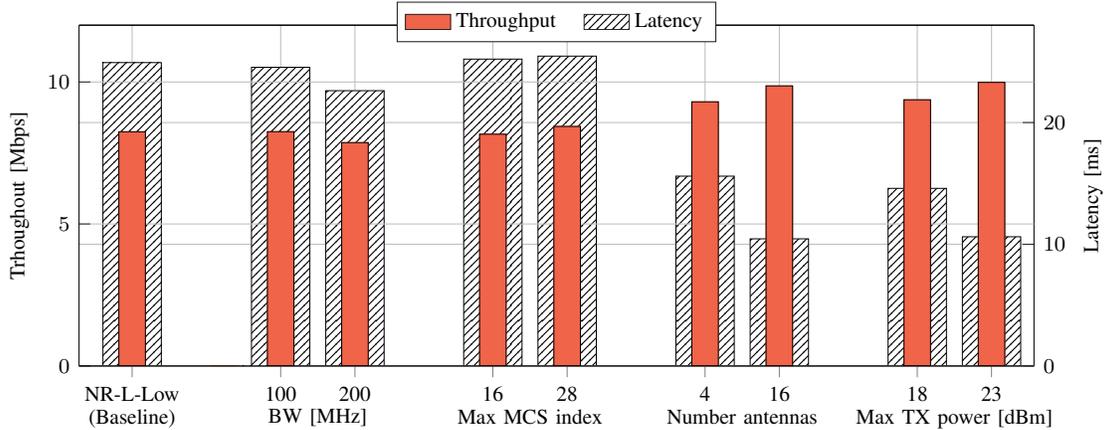

\setlength{\tabcolsep}{0.3em}
\begin{table*}[t!]
  \caption{Numerical results for both InF-DH and InF-SH scenarios, considering both video-stream and data-sensor traffic.
    The latency is computed as the ratio between the number of packets received within the NR-Light requirement set for video-stream (50 ms) and data-sensor (20 ms) devices.}
  \label{Tab:results}
  \centering
  \scriptsize
  \begin{tabular}{c|c|ccccccc}
    InF scenario            & NR-Light configuration     & PRR video & PRR data & Latency video & Latency data & SINR [dB] & Power video [mW] & Power data [mW] \\ \hline
    \multirow{9}{*}{InF-DH} & NR-L-Low (Baseline)        & 0.85      & 0.999    & 0.71          & 0.91         & 14.43     & 48.50            & 1.94            \\
                            & BW: 100.0 MHz              & 0.84      & 0.999    & 0.81          & 0.92         & 11.62     & 38.05            & 1.67            \\
                            & BW: 200.0 MHz              & 0.80      & 0.999    & 0.81          & 0.88         & 7.20      & 34.06            & 1.50            \\
                            & Maximum MCS index: 16      & 0.83      & 1.00     & 0.79          & 0.94         & 12.65     & 40.55            & 1.68            \\
                            & Maximum MCS index: 28      & 0.86      & 0.999    & 0.79          & 0.94         & 11.33     & 38.41            & 1.59            \\
                            & Maximum TX power: 18.0 dBm & 0.95      & 1.00     & 0.81          & 0.95         & 19.34     & 148.47           & 4.84            \\
                            & Maximum TX power: 23.0 dBm & 1.00      & 1.00     & 0.96          & 0.98         & 24.26     & 403.08           & 12.50           \\
                            & UE antenna elements: 4     & 0.94      & 1.00     & 0.86          & 0.95         & 19.60     & 98.94            & 3.41            \\
                            & UE antenna elements: 16    & 0.99      & 1.00     & 0.95          & 0.98         & 26.24     & 286.93          & 9.02            \\ \hline
    \multirow{9}{*}{InF-SH} & NR-L-Low  (Baseline)       & 0.96      & 1.00     & 0.83          & 0.92         & 16.84     & 48.62            & 1.83            \\
                            & BW: 100.0 MHz              & 0.95      & 1.00     & 0.89          & 0.94         & 13.81     & 39.62            & 1.54            \\
                            & BW: 200.0 MHz              & 0.92      & 0.999    & 0.80          & 0.87         & 8.87      & 35.28            & 1.53            \\
                            & Maximum MCS index: 16      & 0.95      & 0.999    & 0.85          & 0.95         & 15.04     & 37.38            & 1.58            \\
                            & Maximum MCS index: 28      & 0.95      & 0.999    & 0.84          & 0.96         & 13.16     & 35.85            & 1.36            \\
                            & Maximum TX power: 18.0 dBm & 1.00      & 1.00     & 0.93          & 0.97         & 23.04     & 132.10           & 4.65            \\
                            & Maximum TX power: 23.0 dBm & 1.00      & 1.00     & 0.96          & 0.99         & 27.45     & 390.62           & 12.36           \\
                            & UE antenna elements: 4     & 1.00      & 1.00     & 0.90          & 0.97         & 23.54     & 92.69            & 3.24            \\
                            & UE antenna elements: 16    & 1.00      & 1.00     & 0.98          & 0.99         & 29.79     & 279.73           & 8.80            \\ \hline
  \end{tabular}
\end{table*}

\subsection{Power Saving Functionalities}
NR-Light communication may follow different transmission patterns compared to, e.g., NR cellular networks. For instance, for an \gls{iiot} scenario, we reasonably expect the traffic to be characterized by short and regular transmissions, e.g., for periodic reporting of sensor data, alternated by long idle periods.
Therefore, the NR-Light system should preemptively specify the slots in which the device has to listen for control messages, and those in which the device can stay idle~\cite{38875}.


Moreover, NR-Light can inherit some of the network functionalities that were specifically designed for low-cost low-power IoT scenarios. In particular, NR-Light can implement a Discontinuous Reception (eDRX) mechanism, a Power Saving Mode (PSM) or wake-up-signals to optimize power consumption during idle modes and increase the battery lifetime.
Additionally, given the directional/beamformed nature of \gls{mmwave} communications, it could be possible to reduce the allowable transmission power budget for devices that are closer to the transmitter, thus promoting reduced interference while increasing the battery~life.


\section{Preliminary Evaluation of an Indoor Factory Scenario using NR-Light at mmWaves}
\label{sec:results}

In this section we evaluate the performance of a reference NR-Light scenario where devices operate at \glspl{mmwave}.
To do so, we performed simulations using the {ns-3} \texttt{mmwave} module,\footnote{Available at \url{https://github.com/nyuwireless-unipd/ns3-mmwave}.} an open-source simulator for 5G \gls{mmwave} networks.
Based on current industry trends and research interests, we focus on a ``factory of the future'' use case (as described in Sec.~\ref{sub:foth}) and consider an indoor factory environment with an area of 20$\times$20 m and 20 static NR-Light devices placed at random positions.
A single base station operating at 28~GHz is deployed on the ceiling, at the center of the area.
The channel model is based on the 3GPP \gls{inf} model, which features two propagation scenarios with different densities of obstacles, i.e., \gls{inf}-SH (sparse) and \gls{inf}-DH (dense).
The \glspl{ue} transmit data to a remote server. Specifically, 90\% of the devices generate \gls{ftp} traffic modeled as a Poisson process of mean 500~kbps (``data-sensor'' devices), while the remaining 10\% (``video-stream'' devices) generate a constant bit-rate video stream of 10~Mbps.

We compared different reduction/simplification strategies for NR-Light, based on those presented in Sec.~\ref{sec:tech}, against a full-blown 5G NR Release 15 device. Specifically, we focus on the impact of bandwidth reduction, relaxation of the maximum \gls{mcs} order and transmission power, and MIMO antenna design at the device.

\subsection{Impact of NR-Light Simplifications}

First, we consider a baseline  NR-Light device (referred to as ``NR-L-Low'') featuring the most drastic reductions, as proposed by the 3GPP~\cite{38875}: i.e., a bandwidth of 50 MHz, a maximum modulation order of 2, one antenna element, and a maximum transmission power of 13 dBm.
Then, we evaluate the impact of progressively relaxing these reductions. 
The performance in terms of throughput and latency is shown in Fig.~\ref{fig:perf-barplot}, while Tab.~\ref{Tab:results} reports numerical results also in terms of \gls{prr} and \gls{sinr}. For better comparison, we further estimate the average energy consumed by the NR-Light devices using the power consumption model in~\cite{desset2020power}.

\subsubsection{Bandwidth reduction}

A progressive increase of the system bandwidth from 50 to 100 and 200~MHz brings a small performance degradation in terms of throughput.
In fact, although a higher bandwidth introduces the possibility of encoding bigger \glspl{tb},
it also causes a higher noise power.
For low-capability devices in a challenging propagation environment, the latter is the dominant effect.

\subsubsection{Maximum modulation order}
We considered a pessimistic baseline with a maximum modulation order equal to~2. Then, we increased this value to 4 and 6, corresponding to \gls{mcs} indices of 16 and 28, respectively~\cite{38214}.
We can see from Fig.~\ref{fig:perf-barplot} that the impact of this solution on the NR-Light performance is basically negligible (the throughput increases by only $2\%$), meaning that the devices experiencing a good channel can satisfy their throughput requirements even with the minimum modulation order possible. Conversely, the remaining devices experience an \gls{sinr} which is too low for the \gls{amc} mechanism to choose a higher \gls{mcs} index.


\subsubsection{Number of antenna elements at the device}

Based on the discussion in Sec.~\ref{sec:tech_mimo}, we consider analog beamforming for NR-Light devices.
In this setting, the number of physical antennas at the \gls{ue} has a significant impact on the system performance. Indeed, an increase from 1 to 4 and 16 antennas introduces a gradual improvement in both the throughput (up to $+16\%$) and latency (around $-60\%$), thanks to the higher gain achieved by beamforming: the average SINR improves by 12~dB when NR-Light devices operate with 16 antennas.

On the downside, Table~\ref{Tab:results} reports that the average power consumed by video-stream (data-sensor) devices increases from 48.5 (1.9) to 286.9 (9.0) mW when increasing the antenna array size from 1 to 16 in the InF-DH scenario. This is due to the higher number of electronic components (specifically phase shifters, converters and power amplifiers) in the RF chain, which makes the choice of the optimal MIMO configuration for NR-Light non-trivial.

\subsubsection{Maximum transmission power}
While decreasing the transmission power typically leads to performance degradation,
in good \gls{sinr} regimes this may promote energy savings, and mitigate interference.
In this context, we studied the impact of decreasing the maximum transmission power of NR-Light devices from 23 dBm (i.e., the Release 15 NR benchmark~\cite{38211_1}) to 13 dBm.
From Fig.~\ref{fig:perf-barplot} we notice that this solution leads to an undesirable degradation of the throughput by $17\%$ and increase of the latency by $134\%$, respectively, due to the resulting lower average \gls{sinr} experienced by the devices. On the other hand, Table~\ref{Tab:results} illustrates that the energy consumption drops by up to approximately 8 (6) times for video-stream (data-sensor) devices, which is certainly in line with NR-Light's requirements.


\begin{figure}[t!]
  \centering
  \begin{subfigure}[t!]{0.44\textwidth}
    \centering
    \includegraphics[width=0.95\textwidth]{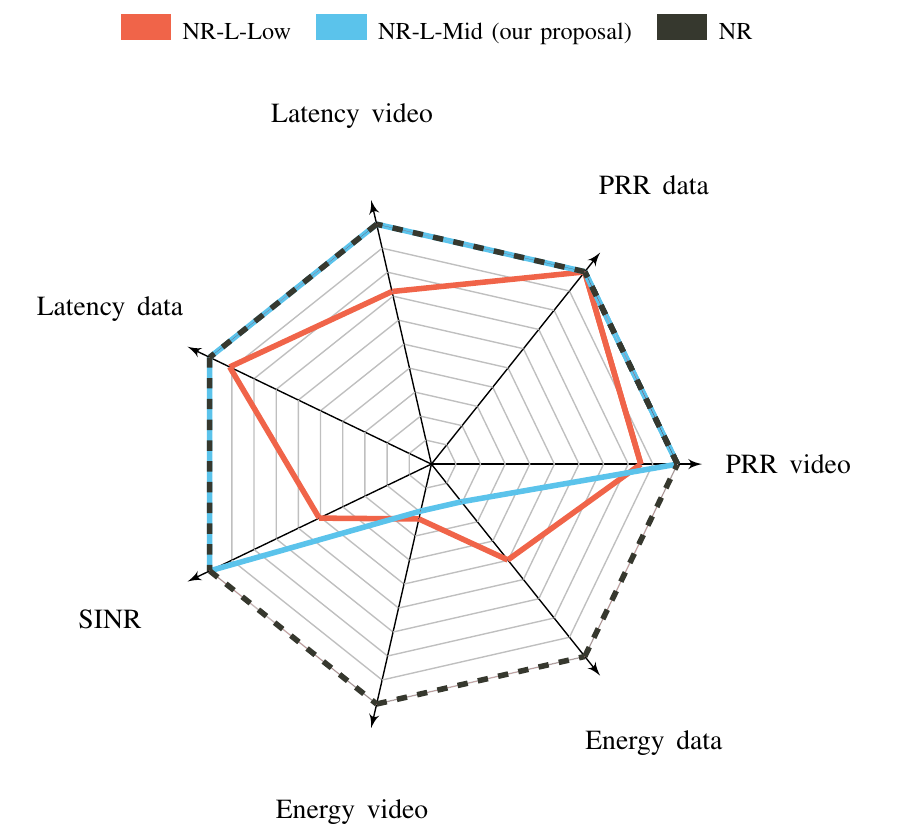}
    \caption{InF-DH scenario.}
    \label{Fig:kiviat-summary-DH}
  \end{subfigure} \\
  \begin{subfigure}[t!]{0.44\textwidth}
    \centering
    \includegraphics[width=0.95\textwidth]{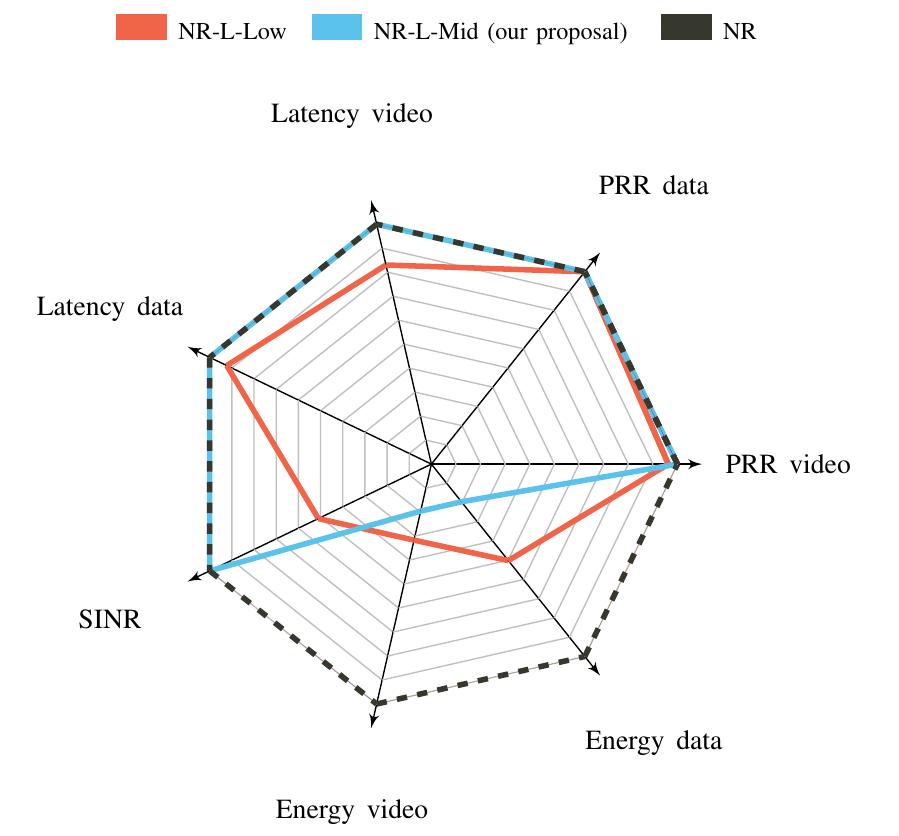}
    \caption{InF-SH scenario.}
    \label{Fig:kiviat-summary-SH}
  \end{subfigure}
  \setlength{\belowcaptionskip}{-0.33cm}
  \caption{Summary of the system performance for the InF scenarios.
    The baseline (NR) is referred to a full-blown NR device featuring a bandwidth of 200 MHz, 16 antenna elements, a maximum transmission power of 23 dBm, and a maximum \gls{mcs} index of 28. For NR-L-Mid (NR-L-Low) devices, we set a bandwidth of 50 MHz, a maximum transmission power of 18 (13) dBm, a maximum \gls{mcs} index of 9, and a number of antenna elements of 4 (1). The values are then normalized with respect to the ``NR'' baseline.}
  \label{fig:kiviat}
\end{figure}

\subsection{Guidelines for an Efficient NR-Light Configuration}

From the simulation results in the previous paragraphs we can draw the following conclusions:
\begin{itemize}
	\item The performance of NR-Light systems is constrained by the \glspl{ue} in the lowest \gls{sinr} regimes;
	\item A system with a single antenna and a maximum transmission power of 13 dBm achieves an insufficient average \gls{sinr}, regardless of the modulation order and bandwidth used, and should be avoided;
	\item Increasing the maximum MCS index would increase the cost/complexity of the devices 
	beyond NR-Light expectations, with limited throughput/latency improvements;
	\item Reducing the bandwidth has generally a positive impact on the devices experiencing bad channel quality.
\end{itemize}


Based on these considerations, we identified the following promising combination of parameters for NR-Light devices, referred to as ``NR-L-Mid" devices, to achieve a good trade-off between network performance and energy consumption:
\begin{itemize}
  \item 50~MHz of bandwidth;
  \item Maximum modulation order of 2 (corresponding to a maximum \gls{mcs} index of 9);
  \item 4 antenna elements;
  \item Maximum transmission power of 18 dBm.
\end{itemize}
We now compare the performance of NR-L-Mid devices with two benchmarks: a typical Release 15 NR device (``NR") and our pessimistic baseline featuring the most extreme NR-Light restrictions (``NR-L-Low") as per the 3GPP~\cite{38875}.

From Fig.~\ref{fig:kiviat}, we see that the proposed NR-L-Mid configuration significantly outperforms the pessimistic NR-L-Low baseline under several metrics, including energy efficiency.
This is due to the fact that, although NR-L-Low devices exhibit the lowest power consumption, implementing the most aggressive restrictions, they transmit for the longest period of time, thus increasing the energy budget.
Moreover, while the NR benchmark is certainly the preferred approach for high-end eMBB/URLLC use cases, NR-L-Mid devices obtain similar end-to-end throughput, latency and \gls{prr} performance, while consuming only 20\% of the energy of a typical NR device.
We claim that the proposed NR-L-Mid configuration strikes the right trade-off between power consumption and portion of time that is spent for data transmission.


\section{Conclusions}
The standardization community agrees that a full-blown 5G NR architecture, while satisfying the needs of high-end eMBB and URLLC applications, may not support mid-market IoT use cases, ranging from video surveillance to industrial automation. To this aim, the 3GPP is working on a new (lighter) version of NR, referred to as NR-Light, to support balanced and mixed IoT-like requirements, ranging among good reliability, acceptable latency and throughput, and low energy/power consumption.
However, while it is foreseen that NR-Light systems operate at mmWaves, numerical results validating this rationale are missing.


In this paper we first illustrate the potential of mmWaves for mid-market IoT use cases.
Then, we highlight possible simplifications of the 5G NR standard to support NR-Light, for example
the use of a lower maximum \gls{mcs} order, a restricted bandwidth support, and simpler antenna configurations. Finally, we demonstrate via simulations the performance of those simplifications, derived from 5G NR, on a typical NR-Light scenario for industrial IoT. 
Our results show that down-selecting features from NR allows for significant gains in complexity and energy consumption, while still meeting application requirements.

\bibliographystyle{IEEEtran}
\footnotesize
\bibliography{ms}

\end{document}